*N. B. Krizhanovskaya, A. A. Krizhanovsky*

*Крижановская Н. Б., Крижановский А. А.*

# SEMI-AUTOMATIC METHODS FOR ADDING WORDS TO THE DICTIONARY OF VEPKAR CORPUS BASED ON INFLECTIONAL RULES EXTRACTED FROM WIKTIONARY

# АВТОМАТИЗИРОВАННЫЕ МЕТОДЫ ПОПОЛНЕНИЯ СЛОВАРЯ КОРПУСА ВЕПКАР НА ОСНОВЕ ПРАВИЛ СЛОВОИЗМЕНЕНИЯ ИЗ ВИКИСЛОВАРЯ [1]

**Abstract.** The article describes a technique for using English Wiktionary inflection tables for generating word forms for Veps verbs and nominals in the Open Corpus of Veps and Karelian languages (http://dictorpus.krc.karelia.ru/). The information concerning Karelian and Veps Wiktionary entries with inflection tables is given. The operating principle of the Wiktionary static and dynamic templates is explained with the use of the *jogi* (river) dictionary entry as an example. The method of constructing the inflection table in the dictionary of the VepKar corpus according to the data of the dynamic template of the English Wiktionary is presented.

**Аннотация.** В статье описан опыт использования данных Английского Викисловаря в построении словоформ вепсских глаголов и именных частей речи в Открытом корпусе вепсского и карельского языков (http://dictorpus.krc.karelia.ru/). Дана справка о наличии словарных статей карельского и вепсского языков и морфологической информации в Английском Викисловаре. Раскрыт механизм работы статических и динамических шаблонов Викисловаря на примере словарной статьи *jogi*. Представлена схема построения таблицы словоизменений в словаре корпуса ВепКар по данным динамического шаблона Английского Викисловаря на примере вепсского существительного *jogi (река)*.

---

[1] The study was supported by the Russian Foundation for Basic Research, grant 18-012-00117.

**Ключевые слова.** Вепсский язык, карельский язык, корпус, словарь, Викисловарь.

## 1. Introduction

The morphological tagging process is one the most laborious works in the corpus linguistics. Large dictionaries with lemmas and word forms are used to perform the morphological tagging.

In the Leipzig Corpora Collection, texts from Internet were crawled and parsed in order to create 400 monolingual dictionaries [Goldhahn et al., 2012]. This is not our case since Veps and Karelian texts are almost absent in the Internet. However the Crúbadán project (Corpus Building for Minority Languages) shows a positive example of automatic search of texts in Internet for under-resourced languages [Scannell, 2007].

The following resources were at our disposal: Wiktionary and traditional dictionaries, native linguists who speak Veps and Karelian, and the programmer who developed the computer system VepKar. The abbreviation VepKar denotes the Open Corpus of Veps and Karelian languages.[2] Researchers at the Karelian Research Centre of RAS have been developing VepKar since 2009 [Zaitseva et al., 2017]. The separate paper [Krizhanovsky et al., 2019] was prepared about Karelian dialects and VepKar corpus in this proceeding.

The article describes a technique for using English Wiktionary inflection tables for generating word forms for Veps verbs and nominals. These generated word forms were added to the VepKar dictionary. The experience of extraction of dynamic templates data from the English Wiktionary is presented in the next section.

---

[2] See http://dictorpus.krc.karelia.ru

**2. Dynamic templates for Veps words and static templates for Karelian in English Wiktionary**

The paper [Metheniti & Neumann, 2018] leads to the idea to extract word forms from English Wiktionary in order to add new lemmas and word forms to the VepKar dictionary. That paper presents the extraction of information from English Wiktionary in 150 languages, but only Veps and Karelian words are of interest for us. Wiktionary entries can contain inflection tables for nouns and verbs generated by using static or dynamic templates.

There is a big difference in an extraction of information from static templates (case of Metheniti & Neumann) and from dynamic templates (our case). The static template (Table 1) is coded in HTML code. It is need to process the HTML code to remove HTML formatting tags and to get word forms from an inflection table.

A completely different thing is a dynamic template, which includes a script (computer program) with inflectional rules. For each language (Veps language in our case) it was a simple matter to write a language-specific parser for dynamic template, but creating a language-independent parser, or filters for 150+ languages will be a major obstacle. This explains why [Metheniti & Neumann, 2018] extracted word forms only from static templates of Wiktionary.

Below we will compare static and dynamic Wiktionary templates and we will describe our approach, where the English Wiktionary served as a donor for the VepKar dictionary expanding.

*2.1. Karelian language and the static template*

There are about 650 Karelian lemmas in the English Wiktionary, but only 30 entries contain inflected forms, therefore no new lemmas with word forms were added to the Karelian dictionary of the corpus VepKar. There is only one template {{krl-decl}} for Karelian words in the English Wiktionary. Note that templates in wiki markup are indicated by double curly braces {{...}}. This static pattern {{krl-decl}} used to show declension tables for 30 Karelian nominals[3].

### 2.2. Veps language and dynamic templates

The situation with Veps words is different in English Wiktionary. There are exactly two dynamic templates: the template {{vep-decl-stems}} for nominals and the template {{vep-conj-stems}} for verbs.

There are 2000 Veps lemmas with about 1000 usages of the inflection-table template {{vep-decl-stems}} in the English Wiktionary. This is a dynamic template that calls a module with Lua programming code. It is worth trying to understand the program in the Lua programming language in order to write the same morphological rules in PHP programming language in VepKar and to get word forms from these 1000 inflection-tables for Veps nouns and adjectives.

### 2.3. Comparison of static and dynamic Wiktionary templates

The *jogi* [4] *(river)* entry in the English Wiktionary will show the difference between dynamic and static templates. This fact, that the word *jogi* exists in both the Karelian and Veps and that there are two types of templates in the Wiktionary entry *jogi*, makes it suitable as an example.

There are different sets of grammatical cases in Veps and Karelian languages (Fig. 1). The inflection tables generated via these templates (Fig. 1) do not show the difference between the static and dynamic templates… And this is correct, since the difference lies not in the tables themselves, but in the ways they are generated. The inflection table construction methods are presented in the Table 1.

---

[3] See pages that link to "Template:krl-decl" at w.wiki/3pb
[4] See https://en.wiktionary.org/wiki/jogi

| Declension of *jogi* | | |
|---|---|---|
| | singular | plural |
| nominative | jogi | jovet |
| genitive | joven | jogiloin |
| partitive | jogie | jogiloi |
| accusative | joven | jovet |
| inessive | joves | jogilois |
| elative | jovespäi | jogiloispäi |
| illative | jogih | jogiloih |
| adessive | jovel | jogiloil |
| ablative | jovelpäi | jogiloilpäi |
| allative | jovele | jogiloile |
| essive | jovennu | jogiloinnu |
| translative | jovekse | jogiloikse |
| instructive | – | jogiloin |
| abessive | jovettah | jogiloittah |
| comitative | jovenke | jogiloinke |
| prolative | joveči | jogiloiči |

| Inflection of *jogi* | | |
|---|---|---|
| | singular | plural |
| nominative | jogi | joged |
| accusative | jogen | joged |
| genitive | jogen | jogiden |
| partitive | joged | jogid |
| essive-instructive | jogen | jogin |
| translative | jogeks | jogikš |
| inessive | joges | jogiš |
| elative | jogespäi | jogišpäi |
| illative | ? | jogihe |
| adessive | jogel | jogil |
| ablative | jogelpäi | jogilpäi |
| allative | jogele | jogile |
| abessive | jogeta | jogita |
| comitative | jogenke | jogidenke |
| prolative | jogedme | jogidme |
| approximative I | jogenno | jogidenno |
| approximative II | jogennoks | jogidennoks |
| egressive | jogennopäi | jogidennopäi |
| terminative I | ? | jogihesai |
| terminative II | jogelesai | jogilesai |
| terminative III | jogessai | — |
| additive I | ? | jogihepäi |
| additive II | jogelepäi | jogilepäi |

*Fig. 1a*. The inflection table of the Karelian noun *jogi* (*river*) generated by the static template {{krl-decl}}

*Fig. 1b*. The inflection table of the Veps noun *jogi* (*river*) generated by the dynamic template {{vep-decl-stems}}

By Wiktionary convention[5], tables that show the forms of nouns are placed in a "Declension" section, while tables that show the forms of verbs are placed in a "Conjugation" section. Fig. 1a corresponds to the convention, but fig. 1b violates it. Why? "This happens because dictionaries are typically the product of several lexicographers' efforts and is constructed, revised, and updated over many years, inconsistencies… necessarily evolve" [Ide & Véronis, 1994]. English Wiktionary has 1600 active editors now[6].

---

5  See Inflection-table templates conventions at w.wiki/3o4

*Table 1.* Static and dynamic templates
used in the Wiktionary entry *jogi (river)*

| N | Karelian language (static template) | Veps language (dynamic template) |
|---|---|---|
| 1 | Wiki page source code (wiki markup) ||
| 2 | ====Declension====<br>{{krl-decl\|title=jogi\|jogi\|joven\|jogie\|joven\|joves\|jovespäi\|jogih\|jovel\|jovelpäi\|jovele\|jovennu\|jovekse\|–\|jovettah\|jovenke\|joveči\|jovet\|jogiloin\|jogiloi\|jovet\|...}} | ====Inflection====<br>{{vep-decl-stems\|jog\|i\|en\|ed\|id}} |
| 3 | Explanation of wiki markup ||
| 4 | The source code contains the *static template* {{krl-decl}} with an explicit listing of all 31 word forms, the template generates an inflection table of Karelian word forms. | The source code contains the *dynamic template* {{vep-decl-stems}} and only 5 template arguments (*jog, i, en, ed, id*). This template calls the Lua script for generating a table with 42 Veps word forms. |
| 5 | Generated inflection tables (see *Fig. 1a* and *Fig. 1b*) ||

---

6 See English Wiktionary statistics at w.wiki/44E

### 2.4. An example of using the data of a dynamic template to add word forms to the Veps dictionary of the VepKar corpus

The template {{vep-decl-stems}} [7] calls the Wiktionary module "vep-nominals" [8]. The inflectional rules for the word form generation were extracted from this code in the Lua programming language and were coded in the PHP language in the VepKar system.

The editor manually copies a text with template name and parameters from the source code of Wiktionary entry (see Table 1, line 2). Then the editor inserts this text into the VepKar corpus to generate word forms (Fig. 2).

The string, which calls the dynamic template {{vep-decl-stems}} with 5 parameters (the stem *jog* and the four endings *i, en, ed, id*), is taken from the English Wiktionary in edit mode (dot-and-dash frame in the bottom of Fig. 2). This string is copied to the *Lemma* field in the VepKar website (dotted frame at the top of Fig. 2) to generate the inflection table of the Veps noun *jogi* (*river*). The generated inflection table in the VepKar corpus for the Veps noun *jogi* is available at: http://dictorpus.krc.karelia.ru/en/dict/lemma/858.

The rules for the word form generation in VepKar made it possible to add 42 word forms to the nominals (based on rules in the module "vep-nominals" [7] in Wiktionary), and 46 verb word forms at once (the template "vep-conj-stems"[9] and the module "vep-verbs"[10]). Thanks to this improvement, technical workers who do not speak Veps have significantly expanded Veps dictionary.

A text with template name and parameters was copied into the VepKar corpus manually (Fig. 2), in order to provide an additional control and verification during the creation of word forms. Since some lemmas in the system already existed, some lemmas had word forms

---

[7] See https://en.wiktionary.org/wiki/Template:vep-decl-stems
[8] See https://en.wiktionary.org/wiki/Module:vep-nominals
[9] See https://en.wiktionary.org/wiki/Template:vep-conj-stems
[10] See https://en.wiktionary.org/wiki/Module:vep-verbs

without grammatical information. Thus, it was necessary to check and remove duplicates.

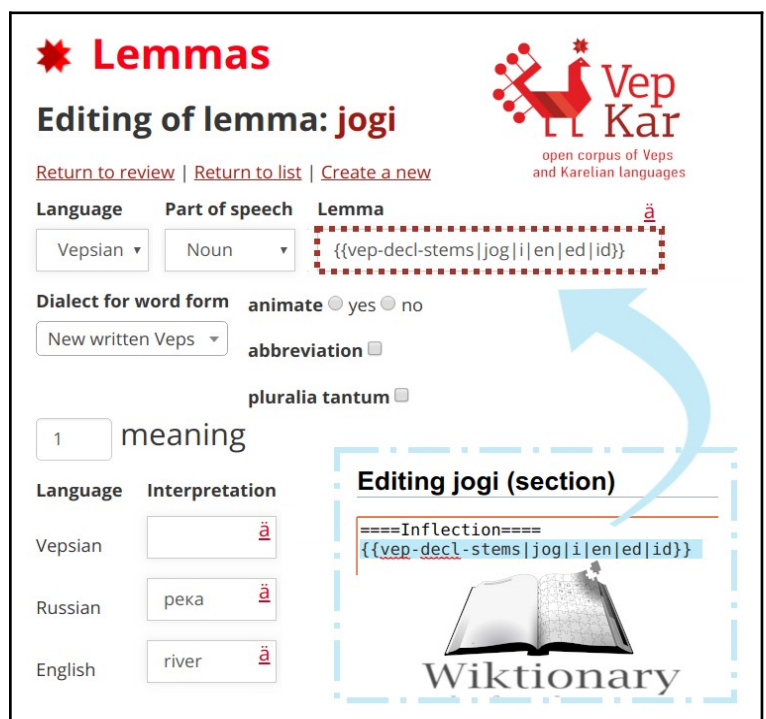

*Fig. 2*. The method of constructing the inflection table in the dictionary of the VepKar corpus according to the dynamic template {{vep-decl-stems}} data of the English Wiktionary using the Veps noun *jogi* as an example

### 3. Discussion and conclusion

Wiktionary inflection tables were used to expand the Veps dictionary of the VepKar corpus. There are a number of reasons for choosing this approach.

- At the first stage, this approach does not require the participation of linguists.
- Rules for constructing inflection tables have already been developed by Wiktionary editors in Lua programming language. It was necessary to adapt these rules to our VepKar corpus system.

After programming inflectional rules in VepKar, the following procedure was applied.

1) Arguments of Wiktionary templates were manually copied into the VepKar dictionary editor, then the VepKar system generated about 40 word forms for each lemma.

2) Then, these rules, initially encoded in a Wiktionary dynamic template, were presented in natural language in the form of a table for discussion with linguists. These rules have been improved and corrected by our linguists.

This table with the rules was a push (and an example) for speeding up work on other Karelian dialects, since it is difficult for linguists to produce a formalized morphological model on their own. A computer program that generates word forms according to the rules is convenient for linguists, since the linguist sees the result and can correct the rule or create a new rule.

Generating word form rules for those grammatical categories that are absent in Wiktionary are going to be refined by Veps linguists in collaboration with programmers in the future. This applies, for example, the illative case for the nominals and all Veps verbs analytical forms.

### References

1. *Goldhahn D., Eckart T., Quasthoff U.* (2012), Building Large Monolingual Dictionaries at the Leipzig Corpora Collection: From 100 to 200 Languages // LREC, Istanbul, Turkey, Vol. 29, pp. 759–765.

2. Ide N., Véronis, J. (1994). Machine Readable Dictionaries: What have we learned, where do we go // Future of Lexical Research, Beijing, China, pp. 137-146.

2. *Krizhanovsky A. A., Krizhanovskaya N. B., Novak I. P.* (2019), Predstavleniye dialektov v Otkrytom korpuse vepsskogo i karel'skogo yazykov (VepKar) [Presentation of dialects in the Open corpus of Veps and Karelian languages (VepKar)] // International scientific conference "Corpus linguistics". Saint Petersburg, 2019.

3. *Metheniti E., Neumann G.* (2018), Wikinflection: massive semi-supervised generation of Multilingual inflectional corpus from Wiktionary // Proceedings of the 17th International Workshop on Treebanks and Linguistic Theories (TLT 2018), December 13–14, 2018, Oslo University, Norway, (155), pp. 147-161.

4. *Scannell K. P.* (2007), The Crúbadán Project: Corpus building for under-resourced languages // Building and Exploring Web Corpora: Web as Corpus Workshop, Vol. 4, pp. 5–15.

5. *Zaitseva N.G., Krizhanovsky A.A., Krizhanovskaya N.B., Pellinen N.A., Rodionova A.P.* (2017), Otkrytyy korpus vepsskogo i karel'skogo yazykov (VepKar): predvaritel'nyy otbor materialov i slovarnaya chast' sistemy [Open corpus of Veps and Karelian languages (VepKar): preliminary selection of materials and dictionary of the system] // International scientific conference "Corpus linguistics". Saint Petersburg, pp. 172-177.

**Krizhanovskaya Natalia (*nataly.krizhanovsky@gmail.com*)**
**Krizhanovsky Andrew (*andrew.krizhanovsky@gmail.com*)**
Institute of Applied Mathematical Research of the Karelian Research Centre of the Russian Academy of Sciences (Russia)